\documentclass[twocolumn]{article}

\usepackage[utf8]{inputenc}
\usepackage{amsmath}
\usepackage{graphicx}
\usepackage{color,soul}
\usepackage[margin=20mm]{geometry}
\usepackage{authblk}
\usepackage{dblfloatfix}
\usepackage{vruler}

\begin{document}

\title{Acceleration of relativistic beams using laser-generated terahertz pulses}

\author[1,2]{Morgan T. Hibberd}
\author[1,3]{Alisa L. Healy}
\author[1,4]{Daniel S. Lake}
\author[1,2]{Vasileios Georgiadis}
\author[1,2]{Elliott J. H. Smith}
\author[1,4]{Oliver J. Finlay}
\author[1,5]{Thomas H. Pacey}
\author[1,5]{James K. Jones}
\author[1,5]{Yuri Saveliev}
\author[1,5]{David A. Walsh}
\author[1,5]{Edward W. Snedden}
\author[1,2]{Robert B. Appleby}
\author[1,3]{Graeme Burt}
\author[1,2]{Darren M. Graham}
\author[1,4]{Steven P. Jamison\thanks{s.jamison@lancaster.ac.uk}}

\affil[1]{The Cockcroft Institute, Sci-Tech Daresbury, Keckwick Lane, Daresbury, Warrington WA4 4AD, UK.}
\affil[2]{School of Physics and Astronomy \& Photon Science Institute, The University of Manchester, Oxford Road, Manchester M13 9PL, UK.}
\affil[3]{Department of Engineering, Lancaster University, Bailrigg, Lancaster LA1 4YW, UK.}
\affil[4]{Department of Physics, Lancaster University, Bailrigg, Lancaster LA1 4YB, UK.}
\affil[5]{Accelerator Science and Technology Centre, Science and Technology Facilities Council, Sci-Tech Daresbury, Keckwick Lane, Daresbury, Warrington WA4 4AD, UK.}

\date{}
\maketitle

Laser-driven acceleration in dielectric structures is a well-established approach \cite{Peralta2013,Breuer2013,Naranjo2012,Niedermayer2018} that may hold the key to overcoming the technological limitations of conventional particle accelerators. However, injecting sub-femtosecond particle bunches into optical-frequency accelerating structures to achieve whole-bunch acceleration remains a significant challenge. A promising solution is to down-convert the laser excitation into the terahertz (THz) frequency regime, where THz pulses with electric fields exceeding 1\,GV/m \cite{Liao2019} have recently been reported. With experimental demonstrations of THz-driven acceleration, compression and streaking with low-energy (sub-100\,keV) electron beams  \cite{Nanni2015,Kealhofer2016,Huang2016,Zhang2018}, operation at relativistic beam energies is now essential to realize the full potential of THz-driven structures. Here we present the first THz-driven acceleration of a relativistic 35\,MeV electron beam at the CLARA research facility at Daresbury Laboratory, exploiting the collinear excitation of a dielectric-lined waveguide (DLW) driven by the longitudinal electric field component of polarization-tailored, narrowband THz pulses. These results pave the way to unprecedented control over relativistic electron beams, providing bunch compression for ultrafast electron diffraction, energy manipulation for bunch diagnostics, and ultimately delivering high-field gradients for compact THz-driven particle acceleration. 

The need to overcome the electrical breakdown threshold currently limiting the achievable accelerating field gradients of radio-frequency (RF)-based accelerators has led to the development of both laser-driven \cite{England2014} and beam-driven \cite{OShea2016} acceleration schemes. With the laser-based schemes also capable of exploiting the femtosecond timing synchronization to laser-generated electron bunches, this has resulted in developments in the optical-infrared regime exploring free-space acceleration \cite{Carbajo2016,Thevenet2016}, laser-driven plasma wakefield acceleration \cite{Faure2004,Leemans2014,Guenot2017} and dielectric laser acceleration (DLA) in scalable, phase-matched dielectric microstructures \cite{Peralta2013,Breuer2013,Naranjo2012,Niedermayer2018}. DLA schemes face significant challenges using sub-micron optical wavelengths to drive the structures, which put extreme tolerances on fabrication and bunch timing jitter, while also limiting the amount of bunch charge that can be supported.

Laser-generated THz radiation exists in the ideal millimeter-scale wavelength regime, making structure fabrication simpler but most importantly providing picosecond pulse cycle lengths well-suited for flexible manipulation of sub-picosecond electron bunches with pC-level charge. In recent years, numerous THz-electron interactions have been explored, including acceleration using a dielectric-lined waveguide \cite{Nanni2015}, electron emission off metal nanotips \cite{Li2016}, streaking and bunch compression both with metallic resonators \cite{Kealhofer2016} and by driving dielectric tubes with circularly polarized THz pulses \cite{Zhao2019}, and the development of novel phase-matching schemes for increased interaction length based on segmented waveguides \cite{Zhang2018}, near-field travelling waves \cite{Walsh2017} and inverse free-electron laser (IFEL) coupling \cite{Curry2018}. The wide scope of work demonstrates the drive towards understanding and exploiting the unique capabilities of THz pulses through structure-mediated interactions for unparalleled control over the spatial, temporal and energy properties of electron beams.

\begin{figure*}[t]
\includegraphics[width=\textwidth,keepaspectratio]{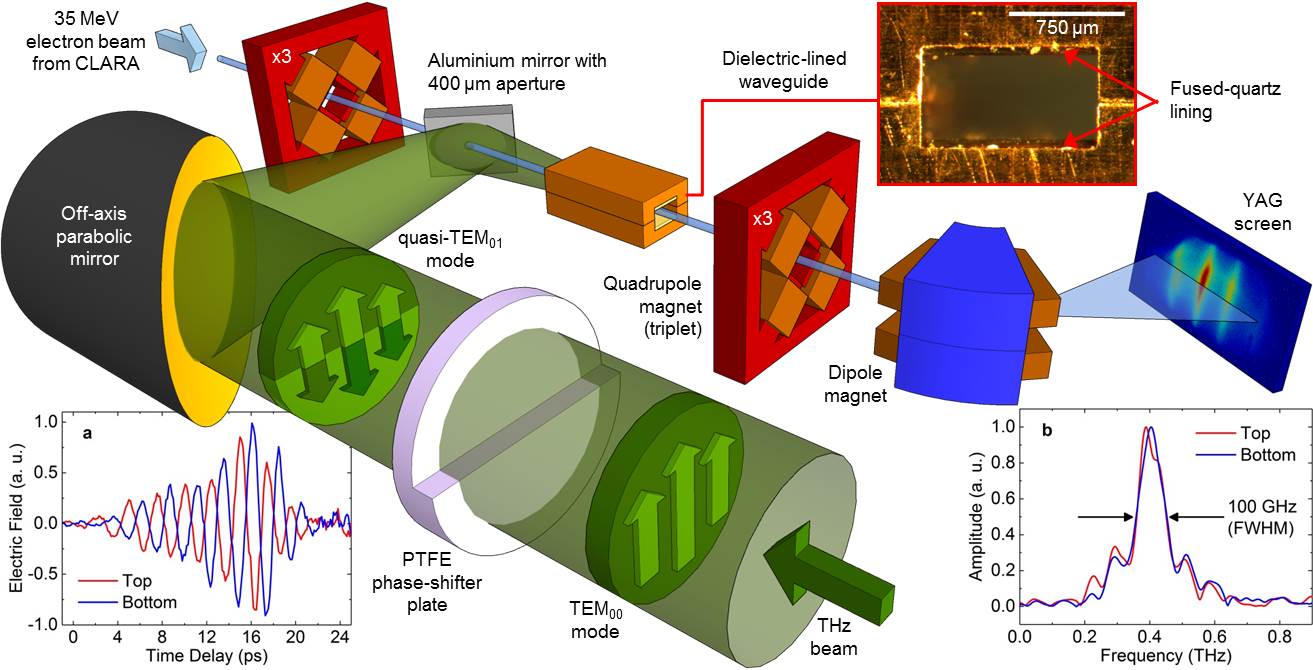}
\caption{\textbf{Experimental setup.} Schematic diagram showing the THz beam converted into a quasi-TEM$_{01}$ mode by a PTFE phase-shifter plate and focused into the DLW for collinear interaction with the 35\,MeV electron beam. A microscope image of the DLW exit is shown, revealing the dielectric lining along the top and bottom surface. Inset: \textbf{a,b,} Electro-optic sampling measurements of the \textbf{(a)} temporal and \textbf{(b)} spectral profiles of the THz pulse transmitted through the top and bottom half of the phase-shifter plate, recorded at the entrance to the DLW coupling horn.}
\label{fig:Setup}
\end{figure*}

Here we demonstrate purely longitudinal, linear acceleration of relativistic electron beams driven by laser-generated THz radiation. In contrast to acceleration of sub-relativistic beams, or transversely-coupled IFEL-driven acceleration \cite{Curry2018}, the resonance frequency for the interaction and the acceleration efficiency remain constant with electron energy gain, offering a viable route to future large-scale, high-energy accelerators.

Narrowband and frequency-tunable THz pulses were phase-velocity matched to 35\,MeV electron bunches (\mbox{$v_{\rm p} = v_{\rm e} =  0.9999c$}) using a longitudinal accelerating mode in a DLW structure for collinear interaction, with an accelerating gradient of 2\,MV/m achieved using modest THz pulse energies. The resulting peak THz-driven acceleration of long-duration (6\,ps FWHM) bunches was determined from the modulation of the energy spectrum, which also revealed quantitatively the time-energy phase-space of the electron bunch. Electron bunches with short duration (2\,ps FWHM) comparable to the period of the THz pulse, are shown to undergo preferential acceleration or deceleration dependent on the timing of electron bunch injection relative to the phase of the THz pulse, demonstrating the route to whole-bunch acceleration of sub-picosecond relativistic particle beams.

The THz-electron beam interaction arrangement is shown schematically in Fig.\,\ref{fig:Setup}. Relativistic 35\,MeV electron bunches with configurable bunch duration and chirp were delivered to the experiment by the CLARA accelerator \cite{Clarke2014}. Magnetic quadrupole triplets provided electron bunch transverse control for coupling into the DLW and for optimizing the energy resolution of the dipole spectrometer. Frequency-tunable, narrowband (100\,GHz FWHM) THz pulses with approximately 2\,$\mu$J energy were generated through chirped-pulse beating \cite{Chen2011,Uchida2015} in a lithium niobate crystal (see Methods). A quasi-TEM$_{01}$ mode was generated by a phase-shifter plate providing a $\lambda/2$ shear and an effective polarity inversion \cite{Cliffe2016,Hibberd2019} along the horizontal midline of the THz beam, which is revealed by the electro-optic sampling measurements in Fig.\,\ref{fig:Setup}a. The DLW structure was a rectangular copper waveguide lined at the top and bottom with 60\,$\mu$m-thick fused-quartz, leaving a vertical 575\,$\mu$m-thick free-space aperture for electron beam propagation. An integrated linearly-tapered horn was used to couple the quasi-TEM$_{01}$ mode of the THz beam into the accelerating longitudinal section magnetic (LSM$_{11}$) mode \cite{Healy2018} of the DLW, which was designed for phase-velocity matching with the electron bunch at an operating frequency of 0.4\,THz (see Supplementary Information).

\begin{figure*}[t]
\includegraphics[width=\textwidth,keepaspectratio]{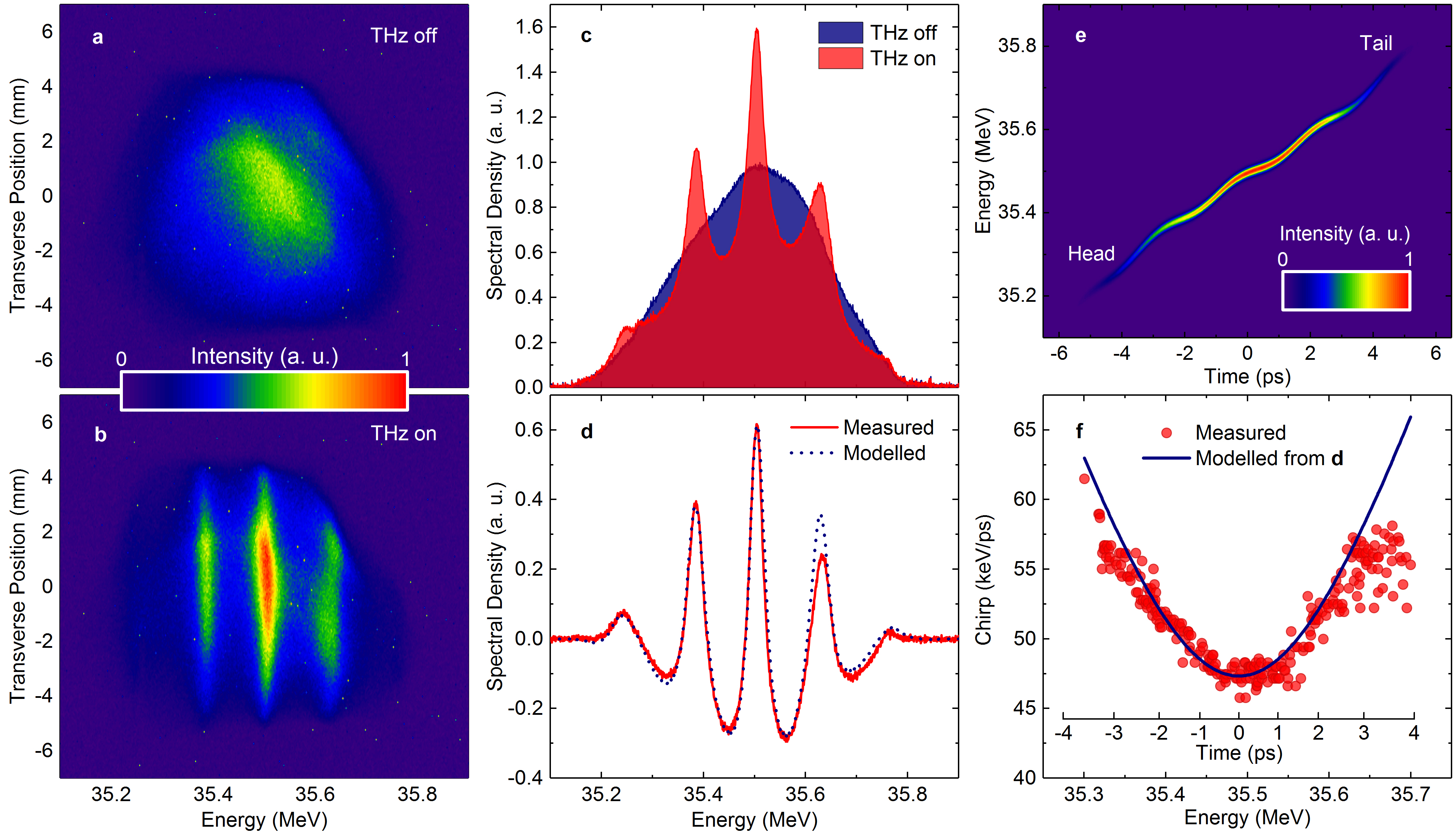}
\caption{\textbf{Multi-cycle energy modulation.} \textbf{a,b,c,} Single-shot images of the long-duration (6\,ps FWHM), high energy spread (330\,keV FWHM) electron bunch with (\textbf{a}) THz off, (\textbf{b}) THz on and (\textbf{c}) the corresponding energy density spectra. \textbf{d,} Measured modulation, extracted from the difference between the THz on and THz off spectra in (\textbf{c}), with optimized model indicating a THz-driven peak electron energy gain of 8.8\,keV. \textbf{e,} Longitudinal phase-space calculation of the modulated electron bunch. \textbf{f,} Measured chirp as a function of energy obtained from 100 successive THz on shots, shown with the modelled chirp, independently obtained from the modelled modulation in (\textbf{d}).}
\label{fig:Modulation}
\end{figure*}

The effect of THz-driven acceleration on an approximately linear chirped, 6\,ps FWHM electron bunch is demonstrated in Fig.\,\ref{fig:Modulation}, with single-shot energy spectra shown for THz off (Fig.\,\ref{fig:Modulation}a) and THz on (Fig.\,\ref{fig:Modulation}b) following temporal overlap of the THz pulse with an electron bunch in the DLW. The difference between the energy spectra in Fig.\,\ref{fig:Modulation}c reveals an energy modulation shown in Fig.\,\ref{fig:Modulation}d, with up to 90\% modulation strength, defined as the peak-to-peak amplitude of the modulation normalized to the amplitude of the unmodulated (THz off) spectrum. The modulation arises from the temporally sinusoidal energy gain induced by the multi-cycle THz pulse on the initial longitudinal phase-space distribution of the electron bunch. With sufficiently large energy gain (and loss), it was possible to impose a significant change in the temporally localized chirp of the bunch, resulting in a non-sinusoidal spectral density. This behavior can be observed from the measured modulation in Fig.\,\ref{fig:Modulation}d, with the peaks occurring where the THz-induced modulation flattened the local chirp to zero. 

The peak THz acceleration ($\delta E_{\rm THz}$) was determined from comparison of the measured single-shot modulation with a calculation of a sinusoidal THz-driven modulation imposed on a model electron bunch. The spatial resolution limitation of the spectrometer, arising from beam emittance and uncorrelated time-slice energy spread served to reduce the observed modulation strength, and were included in the modelling as an effective energy spread ($\delta E_{\rm eff}$) (see Methods). Simultaneous optimization of peak THz acceleration, effective energy spread and chirp (both linear and cubic) provided the modelled modulation shown in Fig.\,\ref{fig:Modulation}d, with a final value of \mbox{$\delta E_{\rm THz}$ = 8.8 keV} obtained. With an effective interaction length limited to approximately 4.3\,mm by group-velocity walk-off and the THz pulse duration (see Supplementary Information), an average acceleration gradient of 2\,MV/m was determined. The model also predicted a value of \mbox{$\delta E_{\rm eff} = 7.9$ keV}, and both linear and cubic chirp components of 47.3\,keV/ps and 0.4\,keV/ps$^3$, respectively.

To further support the above calculation of the acceleration gradient, an independent and direct measurement of the phase-space distribution was obtained from the known THz period \mbox{$\tau = 2.5$\,ps} (0.4\,THz) and the measured energy modulation period $\Delta E(E)$. The experimentally determined chirp of the electron beam ($\Delta E(E)/\tau$) shown in Fig.\,\ref{fig:Modulation}f, was obtained from 100 successive shots using the same electron bunch configuration from \mbox{Fig.\,\ref{fig:Modulation}a-d}. The quadratic dependence indicates the presence of a third-order chirp component, in good agreement with the modelled chirp (solid blue line in Fig.\,\ref{fig:Modulation}f), obtained independently from the single-shot analysis of Fig.\,\ref{fig:Modulation}d. The measured linear chirp (47.5\,keV/ps) was of comparable magnitude with the value (53\,keV/ps) predicted by beam dynamics simulations. However, the degree of phase-space curvature implied by the cubic chirp, revealed from the experimental and modelled data in Fig.\,\ref{fig:Modulation}f, was not expected from the space-charge dominated beam dynamics of the CLARA electron gun. This direct experimental measurement of the time-energy phase-space distribution of a relativistic electron beam is often inaccessible, requiring RF-driven transverse deflecting structures and dedicated electron beam optics \cite{Rohrs2009}. These measurements therefore demonstrate the advantage of THz-driven accelerating structures for both manipulating and exploring the picosecond temporal structure of high-energy electron beams.

\begin{figure}[t]
\includegraphics[width=\columnwidth,keepaspectratio]{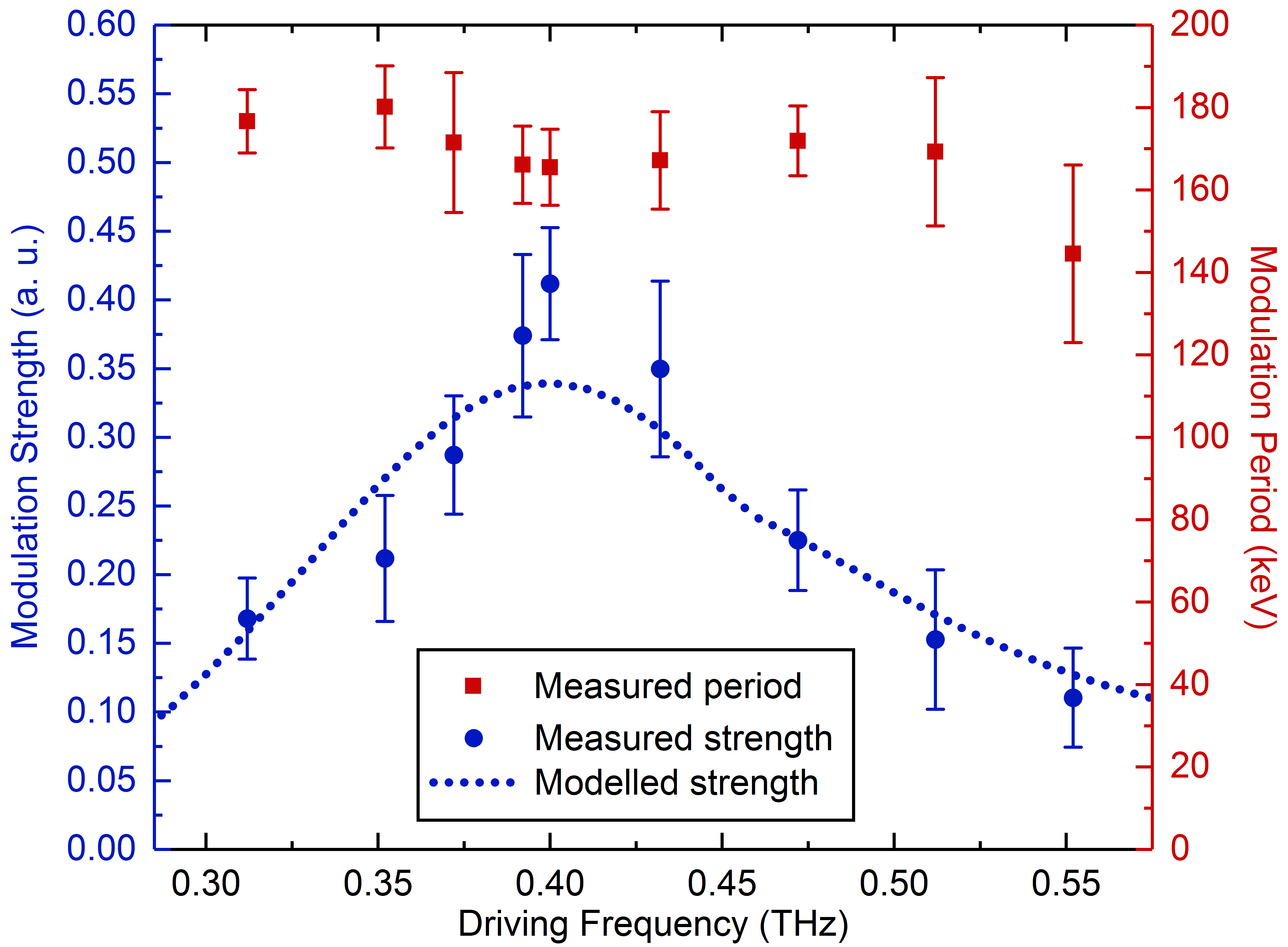}
\caption{\textbf{Phase-velocity matching.} Measurement of both the modulation strength with supporting model obtained from the calculated DLW dispersion, and modulation period, as a function of THz driving frequency. A long-bunch accelerator configuration with increased bunch chirp was used for these measurements.}
\label{fig:Phase_Matching}
\end{figure}

The role of the DLW structure in matching the THz phase-velocity ($v_{\rm p}$) and electron bunch velocity ($v_{\rm e}$) was explored through the frequency-dependent interaction strength, as shown in Fig.\,\ref{fig:Phase_Matching}. The DLW was designed with $v_{\rm p} = v_{\rm e} = 0.9999c$ at 0.40\,THz, and correspondingly the maximum interaction (modulation strength) was observed at this frequency. The length-integrated interaction strength was modelled using the calculated DLW dispersion for a 100\,GHz FWHM bandwidth THz pulse. The width of the resonant frequency response was attributable to both the spectral width of the THz pulse and the temporal walk-off between the THz pulse envelope and the electron bunch. For the 7\,ps FWHM duration THz pulses used here, the group-velocity walk-off limited the effective interaction length to 4.3\,mm (see Supplementary Information). Over the THz frequency range investigated, the modulation period imposed on the bunch was dominated by the interaction efficiency and was observed to remain approximately constant at the value (165\,$\pm$\,10\,keV) obtained at the resonant frequency of 0.4\,THz, as shown in Fig.\,\ref{fig:Phase_Matching}.

\begin{figure*}[t]
\includegraphics[width=\textwidth,keepaspectratio]{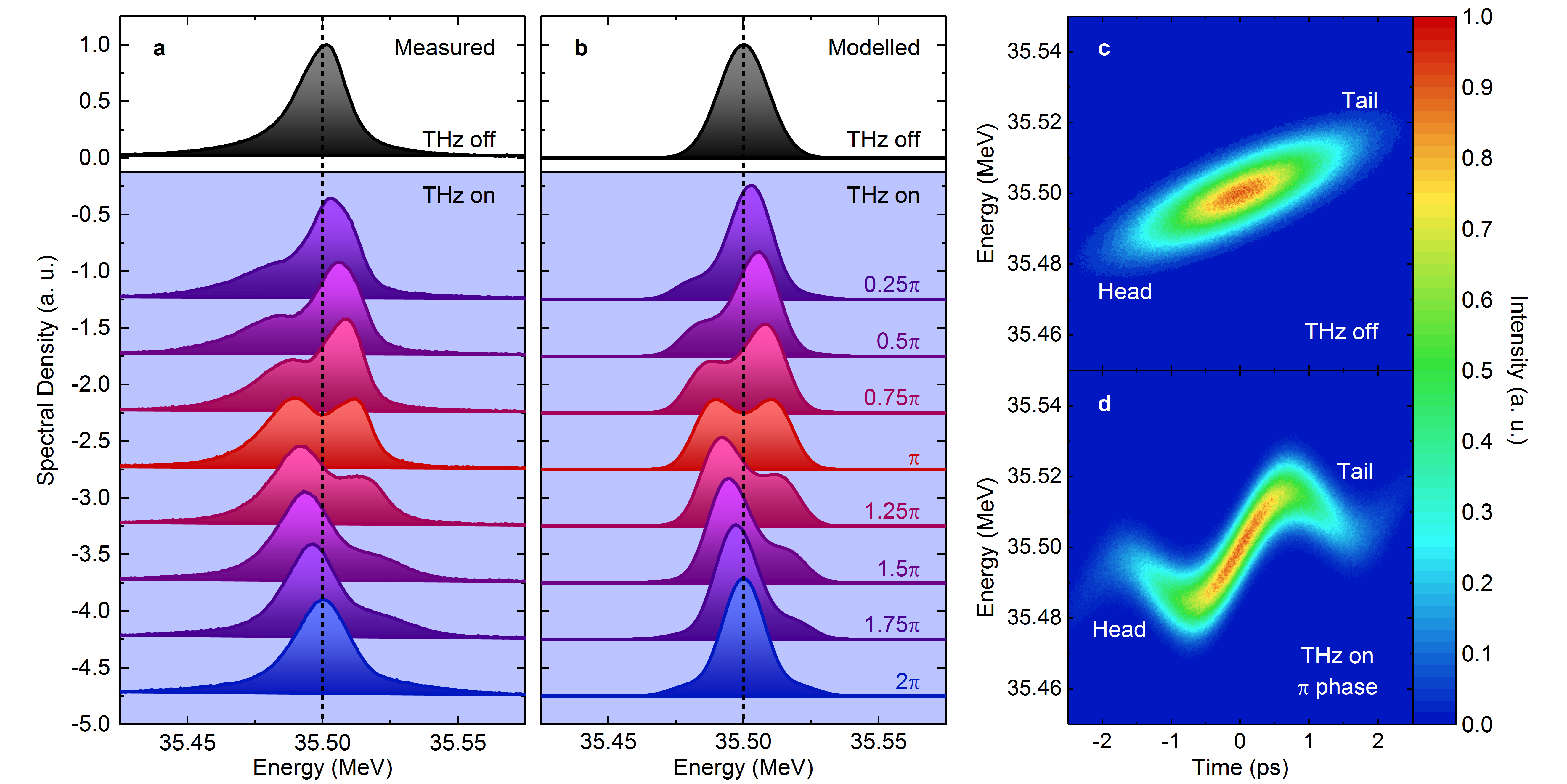}
\caption{\textbf{Bunch energy spreading.} \textbf{a,} Measured single-shot electron energy spectra with THz off and THz on, obtained using short-duration (2\,ps FWHM), low energy spread (\mbox{21\,keV FWHM}) electron bunches. Due to timing jitter, the phase-ordering was determined from the modelled spectra. \textbf{b,} Modelled electron energy spectra shown at discrete bunch injection phases over the full 2$\pi$ range of the THz pulse cycle. \textbf{c,d,} Modelled longitudinal phase-space distribution of the electron bunch with (\textbf{c}) THz off and (\textbf{d}) THz on at $\pi$ injection phase.}
\label{fig:Bunch_Spreading}
\end{figure*}

Compared with optical-frequency DLA structures, THz-driven acceleration offers the potential for injection of an electron bunch into a single half-cycle (or accelerating bucket) of the THz electric field. To demonstrate this, the CLARA accelerator was configured to provide an electron beam with low time-integrated energy spread and an independently measured bunch duration of approximately 2\,ps FWHM (see Methods). With a duration comparable to the THz period, preferential acceleration or deceleration of the bunch was achievable, dependent on the bunch injection timing relative to the phase of the THz pulse. Single-shot electron spectra measured in this configuration are shown in Fig.\,\ref{fig:Bunch_Spreading}a, with observed spectral profiles ranging from symmetric energy spreading to either asymmetric acceleration or deceleration. The maximum measured energy spread FWHM doubled from 21\,keV (THz off) to 42\,keV (THz on) for the symmetric spreading observed at $\pi$ injection phase. This symmetric modulation is similar to that observed in optical-frequency DLA structures \cite{Peralta2013}, where the bunch duration covers many cycles of the optical excitation pulse. However, the asymmetric energy spread observed in Fig.\,\ref{fig:Bunch_Spreading}a was only achievable through a whole-bunch charge-density asymmetry with respect to the THz oscillation period. A model distribution (see Fig.\,\ref{fig:Bunch_Spreading}d) of the low energy spread, short-duration electron bunch was based on particle tracking simulations, with a small residual chirp expected from this accelerator configuration (see Methods). This distribution was then modulated at the THz frequency (see Fig.\,\ref{fig:Bunch_Spreading}e) over the full 2$\pi$ range of injection phases to give the modelled energy spectra in Fig.\,\ref{fig:Bunch_Spreading}b and c. Optimal matching to the measured spectra was achieved using an effective energy spread (see Methods) of \mbox{$\delta E_{\rm eff} = 5.9$ keV}, residual linear chirp of 7.8\,keV/ps and a peak THz-driven acceleration of \mbox{$\delta E_{\rm THz} = 10.0$ keV}, consistent with the value obtained from the multi-cycle modulation analysis in Fig.\,\ref{fig:Modulation}. The experimentally observed and modelled asymmetric acceleration is unambiguous evidence of predominant injection into a single half-cycle of the THz pulse.

In summary, we have successfully demonstrated the linear acceleration of relativistic electron beams using a dielectric-lined waveguide driven by narrowband, frequency-tunable, polarization-tailored THz pulses. We were able to exploit multi-cycle THz-driven modulation to manipulate the electron energy spectra and determine the bunch chirp, while also observe preferential acceleration of bunches with duration comparable to a single cycle of the THz pulse. We demonstrated an accelerating gradient
of 2\,MV/m using modest $\mu$J-level THz
pulses; increasing the gradient by
orders of magnitude will be possible with the implementation of high-energy THz sources \cite{Liao2019}. Our results establish the operation of dielectric structures driven by laser-generated THz pulses in the unexplored relativistic energy regime and pave the way to future acceleration and manipulation of high-energy particles, as well as compact relativistic beam diagnostics.

\section*{Acknowledgements}
We wish to acknowledge the CLARA technical and scientific teams for their support and considerable help on all aspects of the operation of the accelerator. We also wish to acknowledge Peter G. Huggard and Mat Beardsley from Rutherford Appleton Laboratory (RAL)-Space for the manufacture of the dielectric-lined waveguide structure and for the provision of a THz Schottky diode used for THz-electron beam synchronization. This work was supported by the United Kingdom Science and Technology Facilities Council (Grant Nos. ST/N00308X/1, ST/N003063/1 and ST/P002056/1).

\section*{Author contributions}
All  authors  participated  in  the  experiment  and  contributed to data analysis. M.T.H., D.S.L., D.A.W., V.G., O.J.F. and D.M.G. developed the THz source. A.L.H., G.B. and S.P.J. designed the DLW. A.L.H., E.J.H.S., O.J.F., R.B.A. and S.P.J modelled the electron energy spectra and performed the longitudinal phase-space calculations. V.G. characterized the DLW and developed the data acquisition software. T.H.P., J.K.J. and Y.S. analyzed the beam dynamics of the CLARA accelerator. M.T.H., D.M.G. and S.P.J. wrote the manuscript with contributions from all. E.W.S., R.B.A., G.B., D.M.G. and S.P.J. managed the project.

\section*{Competing interests}
The authors declare no competing interests.

\section*{Additional information}
\textbf{Supplementary information} is available for this paper at: https://doi.org/xx.xxxx/xxxxxx.\\
\textbf{Correspondence and requests for materials} should be addressed to S.P.J.

\section*{Methods}

\noindent\textbf{Electron beam.} This experiment was carried out at the CLARA research facility at Daresbury Laboratory. Electron bunches were generated from a copper cathode by 266\,nm, 2\,ps FWHM photoinjector laser pulses at a repetition rate of 10 Hz. The bunches were initially accelerated in a 3\,GHz RF gun to 5\,MeV and then transported to a linac for further acceleration up to 35\,MeV. The relativistic electron bunches were transported to an experimental user station using a magnetic dog-leg, which in combination with the linac, allowed for manipulation of the longitudinal phase-space of the beam. Two main accelerator configurations were used for this experiment; a long bunch with large energy spread (Fig.\,\ref{fig:Modulation}), and a shorter bunch with low energy spread (Fig.\,\ref{fig:Bunch_Spreading}). For the long-bunch configuration, the bunch duration and chirp were determined directly from the THz-driven modulation data (Fig.\,\ref{fig:Modulation} and main text) and were broadly consistent with beam dynamics simulations performed using the code Elegant. For the short-bunch configuration, a bunch duration of 2\,ps FWHM, measured by the method of RF zero-phasing, was used directly in the model analysis of Fig.\,\ref{fig:Bunch_Spreading}. The linear chirp of 7.8\,keV/ps was in line with accelerator tuning and space-charge expectations. In the experimental user station, the electron beam was transversely focused into the DLW by an upstream quadrupole triplet (see Fig.\,\ref{fig:Setup}) to a RMS transverse size of 100\,$\mu$m. After transmission through the DLW a downstream quadrupole triplet was used for matching the beam into the energy spectrometer and minimizing the horizontal $\beta$-function.

\noindent\textbf{Laser systems.} The experiment was performed using a customized terawatt (TW) laser system \cite{Priebe2008}, capable of producing up to 800\,mJ laser pulses at 10\,Hz with a center wavelength of 800\,nm and Fourier-limited pulse duration of 60\,fs. The TW system comprised of a Ti:sapphire oscillator (Micra, Coherent) producing 800\,nm, 30\,fs, 4\,nJ pulses at 83\,MHz, which were stretched to approximately 150\,ps and used to seed a Ti:sapphire regenerative amplifier (Legend, Coherent) providing 800\,nm, 50\,fs (Fourier-limited), 1\,mJ pulses at 1\,kHz. The uncompressed output of the regenerative amplifier was routed through a multi-pass Ti:sapphire amplifier (MPA), where two frequency-doubled Nd:YAG lasers (Powerlite II, Continuum) producing 532\,nm, 10\,ns, 1.5\,J pulses at 10\,Hz provided the pump energy for amplification, which following re-compression to 60\,fs resulted in TW peak powers. For the experiment, a laser pulse energy of approximately 50\,mJ was delivered to the electron beam experimental user station through an evacuated beam transport line.

\noindent\textbf{Terahertz generation.}
Narrowband THz pulses were generated using a chirped-pulse beating scheme combined with tilted pulse-front pumping of a 0.6\% MgO-doped stoichiometric lithium niobate (LiNbO$_3$) crystal. Chirped-pulse beating was achieved by adjusting the TW grating-pair compressor to chirp the input laser pulse and a Michelson interferometer was used to generate THz radiation  at the beat frequency set by the interferometric combination of the two chirped laser pulses in the LiNbO$_3$ crystal. With an input chirped pulse duration of approximately 12\,ps FWHM, THz pulses with centre frequency of 0.40\,THz and bandwidth of 100\,GHz were generated. The tilted pulse-front was achieved using a diffraction grating with groove density of 1700\,mm$^{-1}$, incident angle of 31.2$^{\circ}$ and diffracted angle of 57.5$^{\circ}$. A 4f-lens configuration consisting of two cylindrical lenses of 130\,mm and 75\,mm focal length was used to image the pulse-front tilt in the LiNbO$_3$ crystal. The resulting pump 1/\textit{e}$^2$ spot size on the LiNbO$_3$ crystal was \mbox{5 mm x 10 mm} and with a pump pulse energy of \mbox{18 mJ}, THz pulses with an approximate energy of 2.1\,$\mu$J were measured at the crystal surface using a pyroelectric detector (THZ-I-BNC, Gentec). The THz radiation was collected by a 90$^{\circ}$ off-axis parabolic mirror (OPM) of 152.4\,mm focal length and routed by silver mirrors into a vacuum chamber through a quartz window. Focusing of the THz radiation was achieved using a 228.6\,mm focal length OPM combined with a 100\,mm focal length TPX lens, resulting in a 1/\textit{e}$^2$ spot size of approximately 3\,mm at the waveguide coupler entrance and a measured THz energy of approximately 0.79\,$\mu$J. A small fraction (5\%) of the chirped 12\,ps laser beam was re-compressed back to 60\,fs FWHM by a second grating-pair compressor and focused on to a 500\,$\mu$m-thick, (110)-cut ZnTe crystal in a back-reflection geometry for electro-optic sampling measurements of the THz electric field at the entrance of the waveguide. For generating the quasi-TEM$_{01}$ THz mode, a 40\,mm-diameter polytetrafluoroethylene (PTFE) phase-shifter plate was made with a thickness difference between the top and bottom halves of approximately 800\,$\mu$m.

\noindent\textbf{Synchronization.} The ultrafast 83\,MHz Ti:sapphire oscillator (at the front end of the TW laser system) was synchronized to a reference frequency from the CLARA master clock using a commercial phase-lock loop electronics module (Synchrolock, Coherent). The master clock was derived from the Ti:sapphire oscillator (Element, Spectra-Physics) integrated into the CLARA photoinjector laser system. The RMS timing jitter between the two oscillators was measured to be approximately 100\,fs. Temporal overlap of the THz pulse with the electron bunch at the DLW entrance was achieved using a Schottky diode, which detected both the driving THz pulse directly and the coherent transition radiation (CTR) generated by the 35\,MeV electron beam incident on a vacuum-metal interface inserted into the beam path. Phase adjustments of the commercial phase-lock loop electronics module were used for coarse control, after which a mechanical delay stage (with range of 1\,ns and step-size of 33\,fs) on the THz beam line was used for fine control of the temporal overlap.

\noindent\textbf{Waveguide structure.}
The waveguide design was a hollow, rectangular copper structure lined at the top and bottom with 60\,$\mu$m-thick fused quartz (using a glycol phthalate adhesive), leaving a 575\,$\mu$m-thick free-space aperture of 30\,mm length and 1.2\,mm width for electron beam propagation (see Supplementary Information). To maximize coupling of the THz radiation into the accelerating mode of the DLW, a tapered horn structure was fabricated with an entrance aperture of 3.25\,mm by 3.18\,mm and length of 23\,mm. A 45$^\circ$ aluminium mirror with a 400\,$\mu$m aperture aligned to the DLW was used to spatially overlap the incident reflected THz radiation with the transmitted electron beam. The overall DLW/coupler/mirror structure was located on a motorized 5-axis translation stage providing fine control over the positioning and tilt angle for optimization of the THz and electron beam transmission.

\noindent\textbf{Modelling.}
An effective energy spread ($\delta E_{\rm eff}$) was used in the model analyses to describe the spatial resolution ($\sigma_{\rm rms}$) of the electron spectrometer and was related to the spectrometer dispersion ($D$), uncorrelated time-slice energy spread ($\delta E_{\rm uncorr}$), beam emittance ($\epsilon$) and beam optical $\beta$-function by \mbox{$\delta E_{\rm eff} = \sigma_{\rm rms}/D = \sqrt{\delta E_{\rm uncorr}^2 + \epsilon^2\beta^2/D^2}$}, where $\sigma_{\rm rms}$ was determined to be 270\,$\mu$m and 200\,$\mu$m for the beam conditions used in the measurements of Fig.\,\ref{fig:Modulation} and Fig.\,\ref{fig:Bunch_Spreading}, respectively.

The modelled frequency-dependent interaction strength shown in Fig.\,\ref{fig:Phase_Matching} was calculated using 
\mbox{$\delta U = \int_0^L E_{\rm z}(z,t=(z-z_{0})/v_{\rm e}){\rm d}z$} for particles injected with phase-offset $z_0$. The propagating field $E_{\rm z}(z,t)$ at the particle position was calculated from the frequency domain spectrum of the THz pulse and the frequency-dependent propagation wavevector for the DLW.\\

\noindent\textbf{Data availability.} The data associated with the paper are openly available from the
Mendeley Data Repository at: http://dx.doi.org/xxxxxxxxxxxx.

\end{document}